**Self-Organized Platinum Nanoparticles on Freestanding Graphene**

Peng Xu,[†] Lifeng Dong,*[,‡,§] Mehdi Neek-Amal,[∥,#] Matthew L. Ackerman,[†] Jianhua Yu,[‡] Steven D. Barber,[†] James Kevin Schoelz,[†] Dejun Qi,[†] Fangfang Xu,[⊥] Paul M. Thibado,*[,†] and Francois M. Peeters*[,∥]

[†]Department of Physics, University of Arkansas, Fayetteville, Arkansas 72701, United States

[‡]College of Materials Science and Engineering, Qingdao University of Science and Technology, Qingdao 266042, China

[§]Department of Physics, Astronomy, and Materials Science, Missouri State University, Springfield, Missouri 65897, United States

[∥]Departement Fysica, Universiteit Antwerpen, Groenenborgerlaan 171, B-2020 Antwerpen, Belgium

[#]Department of Physics, Shahid Rajaee Teacher Training University, Lavizan, Tehran 16788, Iran

[⊥]State Key Laboratory of High Performance Ceramics and Superfine Microstructures, Shanghai Institute of Ceramics, Shanghai 200050, China

*Address correspondence to donglifeng@qust.edu.cn, thibado@uark.edu, francois.peeters@uantwerpen.be.

**ABSTRACT**   Freestanding graphene membranes were successfully functionalized with platinum nanoparticles (Pt NPs). High-resolution transmission electron microscopy revealed a homogeneous distribution of single-crystal Pt NPs that tend to exhibit a preferred orientation. Unexpectedly, the NPs were also found to be partially exposed to the vacuum with the top Pt surface raised above the graphene substrate, as deduced from atomic-scale scanning tunneling microscopy images and detailed molecular dynamics simulations. Local strain accumulation during the growth process is thought to be the origin of the NP self-organization. These findings are expected to shape future approaches in developing Pt NP catalysts for fuel cells as well as NP-functionalized graphene based high-performance electronics.

Direct methanol and ethanol fuel cells are excellent power sources due to their high energy density, low pollutant emission, low operating temperature, and the ease of handling liquid fuel.[1] In these devices, oxygen undergoes a reduction facilitated by a catalyst while the fuel is catalytically oxidized at the anode. The catalyst is critical to the reactions, and the most popular candidate for both reduction and oxidation is the noble metal platinum (Pt). However, the worldwide supply of Pt is inadequate for mass production. To combat this problem, scientists have moved toward increasing the Pt surface-area-to-volume ratio through the use of nanoparticles (NPs). Platinum NPs 2 nm to 5 nm in size, and even as small as 0.9 nm, can be used to replace a solid Pt film without losing catalytic activity due to quantum size effects.[2, 3] Among the most effective supports for Pt NPs in electrochemical catalysis are carbon materials due to their large surface areas for dispersion of the NPs, a porous structure for transferring reactants and products, and good electrical conductivity for electrochemical reactions.[4, 5] In fact, the two-dimensional (2D) carbon allotrope graphene may provide the ideal support for Pt NP catalysis, with its unmatched electrical conductivity, strength, and surface-to-volume ratio.[6, 7] Furthermore, some theoretical studies have predicted that by placing certain elements on graphene the catalytic reactions are enhanced.[8]

First-principles studies using Pt NPs have shown that NP can bind differently to a graphene sheet depending on the number of Pt atoms and their geometry.[9] For example, as the number of Pt atoms in the NP increases, the Pt-C interaction energy *per* contacting Pt atom decreases, which results in fewer planar NPs and more 3D clusters.[10] The basic conclusion being that the interfacial interaction between graphene and a Pt NP strongly depends on the geometry, size, and contact details. Furthermore, temperature affects the morphology of suspended graphene, the adsorption process, and binding energy,[11] making this a difficult system to model without experimental data for comparison. As of yet, however, it has not been demonstrated experimentally that a pristine graphene, nor the more complicated case of freestanding graphene, can be functionalized with Pt NPs in the uniform manner required for effective catalysis.

The two leading techniques for characterizing graphene and its derivatives at the atomic scale are transmission electron microscopy (TEM) and scanning tunneling microscopy (STM). These two techniques complement one another by providing information that the other cannot. For example, perturbations caused by the biased STM tip make it difficult to produce high-quality, large-scale STM images of freestanding graphene, but TEM images have been successfully obtained.[12] On the other hand, the damage caused by high-energy electrons makes TEM imaging of suspended graphene impractical at a sub-nanometer scale,[13] while STM has been used to acquire atomic-scale images without damaging the membrane.[14, 15] TEM has additionally been used to obtain atomic-resolution images of crystals and molecules,[16, 17] including particles adsorbed on or encapsulated in graphene.[13, 18-20] In fact, using graphene as a TEM substrate has been shown to enhance the resolution of the resulting nanoparticle images.[18]

Meanwhile, STM can measure the height of NPs with unmatched precision and locally manipulate flexible samples. In tandem, these two techniques provide a complete picture of the NP/graphene system at the atomic scale.

In this work, graphene was first suspended on copper TEM grids and imaged with scanning electron microscopy (SEM) to confirm coverage, as shown in Figure 1a. A zoomed-in view of the marked grid hole is shown in Figure 1b. Close inspection of a large number of images reveals the grid is more than 90% covered with graphene (*i.e.*, the black regions lack graphene). Pt NPs were then grown on freestanding graphene using a single-step sputtering process, in which argon ions ($Ar^+$) bombarded a Pt target, and the ejected Pt landed on the membrane, as illustrated in Figure 1c. The composite films were then imaged using an unprecedented combination of high-resolution TEM (HRTEM) and STM. Insight into the interactions between graphene and the NPs was also obtained by mechanically manipulating the functionalized graphene with the STM tip. Through these various techniques, combined with molecular dynamics (MD) simulations, it was determined that the Pt NPs are uniform in size, they organize into a nearly periodic arrangement on the sample surface, and the graphene does not encapsulate the NPs. Based on the NP size and number density reported below, use of Pt-NP functionalized graphene would result in an 80% reduction in the metal volume compared with a 2 nm thick continuous Pt film, while maintaining a similar effective surface area for catalysis.

**RESULTS AND DISCUSSION**

Bright-field TEM images of the Pt NPs supported by freestanding graphene are displayed in Figure 2. Approximately 2,000 NPs, seen as minute black points, are uniformly dispersed on the 170 nm × 170 nm area shown in Figure 2a. Zooming in, the 30 nm × 30 nm image shown in Figure 2b reveals that the NPs prefer a specific nearest-neighbor distance, tend to line up, and have a number density of $6.6 \times 10^{12}$ cm$^{-2}$. Analysis of all 70 Pt NPs in this view reveals that more than 50% show a clear and similar crystal structure, while the remaining NPs are indeterminable. Zooming further still, a 10 nm × 10 nm image displayed in Figure 2c shows the highest magnification obtained, at which point the atomic-scale details of the NPs can be clearly seen. The measured inter-planar spacing of 0.20 nm and 0.23 nm (marked on Figure 2c) corresponds to those of (200) and (111) planes, respectively, in a Pt single crystal. Fast Fourier transform analysis shows that the zone axis of these NPs is along the $[01\bar{1}]$ direction. A size analysis of the particles is presented in the histogram in Figure 2d, showing a narrow size distribution with an average diameter of 1.4 nm ± 0.2 nm. To decipher the periodicity of the NP arrangement observed in Figure 2b, the autocorrelation function (ACF) of that image was calculated and is displayed in Figure 2e. The overall symmetry is not very well defined, but it does slightly favor the honeycomb symmetry of the graphene lattice. A line profile taken from the ACF along the drawn diagonal line is shown in Figure 2f. From this, the average particle size, as determined by the distance at 1/$e$ of the initial peak, is found to be about 1.3 nm, consistent with the results of the size analysis reported above. The presence of the second peak in the line profile confirms ordering in that direction, while its location of 4 nm is the preferred nearest-neighbor separation of the NPs.

To learn about the height and the top surface topography of the Pt NPs, atomic-scale STM images were obtained. A typical filled-state, 12 nm × 12 nm STM image is shown in Figure 3a, revealing an artificially stretched honeycomb lattice[21] near a pair of large white features (indicating increased height). A white box marks the original location of the lower right inset and highlights the stretched honeycomb lattice, while the left inset was acquired far from the NPs and shows the normal lattice constant. The large white shapes in figure 3a are determined to be Pt NPs based on the number density and average size observed in this and other STM images, combined with the HRTEM data showing the same details throughout the surface. Vertical and horizontal line profiles, running through the center of Figure 3a along the lines shown in red and blue, were extracted from the image. The horizontal line profile is plotted in blue directly below the image, while the vertical line profile is plotted in red at the bottom. A large peak corresponding to the central NP appears in both curves. In the horizontal profile, moving from left to right, we find that the tip height increases approximately 0.6 nm to the top of the NP and decreases 1 nm to the base, while in the vertical profile it increases 1.2 nm to the top and decreases 0.8 nm to the base. The different height changes from one side to the other suggest that the local morphology of graphene is affected by the presence of the Pt NP. Note, the NP widths from the STM line profiles are affected by tip-sample convolution effects,[22] but we can rely on the TEM data for that information. From the TEM data the width of the nanoparticles is about 1.4 nm, and this is larger than its height (from the STM data), indicating a preference for wetting the surface over 3D growth. Furthermore, we can conclude that only 2–3 layers of Pt atoms are present in this particular NP (*i.e.*, Pt lattice constant = 0.39 nm).

Atomic-scale details on the surface of the Pt NP can be observed after a 5 nm × 5 nm area was cropped from Figure 3a and given a compressed color scale to enhance the details as shown in Figure 3b. We are able to faintly resolve atomic rows, which can be seen running diagonally from the lower left to the upper right. A local height line profile extracted along the overlaid line in Figure 3b, shows the details better and is plotted just below the image. Clear oscillations occur with a small height corrugation of only 0.05 nm. The row-like symmetry of the surface topography is inconsistent with the honeycomb structure of graphene, but is consistent with exposed Pt. The average distance between individual peaks (*i.e.*, the individual rows) is ~0.20 nm, which matches the distance between the rows of atoms shown in the TEM image of Figure 2c and suggests we have a $(01\bar{1})$-oriented surface with (200)-oriented planes. A simple drawing of a Pt structure having a $(01\bar{1})$-oriented surface is shown below the line profile. The surface has five rows of Pt atoms (highlighted) running from the lower left to the upper right. All totaled, we can confirm that the Pt NPs are securely affixed to the suspended graphene, as they are stable enough to be imaged with the STM, and that the graphene does not wrap around the NP. This is in contrast to systems where graphene encapsulation has been observed for nanocrystals[23] and bacteria.[24]

In order to better understand the adsorption mechanism, we turn to MD simulations carried out using reactive force fields. (See Methods.) We considered a square-shaped graphene with a computational unit cell of 12 nm × 12 nm which contains 8640 carbon atoms. Pt atoms were placed on the graphene sheet, initially arranged in a square lattice of lattice constant 0.3 nm and with a height of 0.34 nm, as shown in Figure 4a. After MD relaxation (~1 ns) at low temperature (10 K), a new self-assembled structure had developed and is shown from three different viewpoints in Figure 4b (see also the supplemental movie). The Pt atoms form a NP with no well-defined shape but having a lateral width of about 3 nm and a height of 0.5 nm, in quantitative agreement with our experimental results shown in Figure 3. The important discovery here is that one atomic layer of Pt atoms over graphene is energetically unfavorable even at very low temperature, while condensing the NP decreases the total energy of the system. In addition, we found that competition between interatomic forces (*i.e.*, between Pt-Pt and Pt-C atoms), limits the vertical growth to 2–3 layers of Pt atoms. The radial distribution function for Pt atoms in the NP of Figure 4b revealed that almost all nearest neighbor Pt-Pt bonds were 0.278 nm long, and the lattice structure was a compacted structure different from perfect face-centered cubic (fcc). Interestingly, the NP is on average *elevated* above the substrate, with very few Pt atoms bonded to the graphene sheet, as evidenced by the right inset of Figure 4b. The Pt-C bonds, which are mostly 0.2 nm long, are *covalent* in nature and ensure the stability of the NP on the surface up to room temperature, though it detached from the surface beyond that. Additional calculations and simulations were performed to test the Pt NP number density and crystallinity observed with TEM. Here, when five NPs with pre-imposed fcc structure are spaced apart from each other and allowed to relax at 300 K, they are stable, as shown in Figure 4c. This confirms the viability of the NP arrangement seen in Figure 2, assuming thermal fluctuations in the graphene sheet remain small. These NPs are found to sit on top of the graphene surface, and Pt atoms do not flow from one NP to the other. In fact, the fcc NPs themselves maintained their structure even up to 1000 K.

In order to experimentally probe the strong binding between the Pt NPs and the graphene sheet, we used a relatively novel scanning tunneling spectroscopy (STS) technique.[14] While imaging, we interrupted a topography scan already in progress and varied the STM tip bias voltage $V$ from 0.1 V to 3.25 V, measuring the change in tip height $Z$ required to keep the current constant (*i.e.*, feedback on configuration). As the voltage increases, however, so does the electrostatic attraction between the STM tip and the grounded sample, causing the graphene to flex towards the tip. Because the STM feedback circuit is on, the tip simultaneously retracts to maintain a constant tunneling current. This measurement was repeated for three different samples using the same current setpoint of 1.00 nA, and typical results are displayed in Figure 5a. The lowest curve shows almost zero displacement of our inflexible control Au sample, the middle curve is what we measure for pristine freestanding graphene and shows a net displacement of about 30 nm, and the top curve is what we measure for Pt-functionalized freestanding graphene, showing a net displacement of about 140 nm. Both freestanding graphene displacements are several orders of magnitude too large to be explained only by changes in the local density of states, which could be at most on the order of 1–3 nm.[25] Therefore, we are primarily measuring mechanical movement of the suspended graphene membrane.

We can estimate the electrostatic force[26] between the tip and the sample to be a few nano-Newtons. This force is pulling on the NP-graphene system and causing it to move toward the STM tip, and since these displacements are fully reversible and repeatable, we can conclude that this force is not large enough to detach a Pt NP from the freestanding graphene surface. This is consistent with the strong covalent bond predicted by our MD simulations. In addition, with the Pt covalently bonded to the surface, the local bonding of the graphene is altered from $sp^2$ to partial $sp^3$. This causes the surface curvature to be convex and slightly elevates the Pt NP. Furthermore, the local $sp^3$ bonding softens the graphene perpendicular to the surface (*i.e.*, the same direction of the applied electrostatic force) and explains the lower voltage needed to reach a 30 nm displacement for the Pt-functionalized film compared to pristine. The important finding here is that the graphene sheet is easier to distort in the presence of Pt NPs, which we deduced from the larger movement in the STS measurements. Generally, the presence of Pt NPs enhances the roughness of graphene and causes the formation of extra ripples, which results in non-uniform carbon-carbon bonds and a non-uniform strain in the membrane. The latter is due to the partial $sp^3$ hybridization between some Pt atoms and makes the graphene weaker.[27]

Due to the surprising nature of the result, we performed extra MD simulations to further confirm that the graphene surface elevates the Pt NPs rather than wraps around them. We formed a Gaussian-shaped depression in freestanding graphene and filled it with Pt atoms as shown in Figure 5b. After MD relaxation, the initial concave curvature in the graphene surface disappears and even becomes slightly convex. Note that the original concave curvature in graphene is stable in the absence of Pt, so this significant change in the deformation is due to the Pt NP. The height of the Pt atoms is not uniform everywhere, which is consistent with the experimental results shown in Figure 3. The overall self-assembled geometry is due to the preference of forming more covalent bonds between Pt-Pt, rather than Pt-C. Only two or three Pt-C bonds are preferred due to the competition with the elastic energy of graphene. Generally, when weaker *van der Waals* interactions are dominant, the graphene tends to wrap around the NP, while when stronger covalent bands are involved the elevation occurs.

The elevated feature of the Pt NP also helps to explain the self-organization previously discussed in relation to Figure 2. During the deposition process, Pt atoms and possibly small clusters are ejected from the target and land on the graphene membrane. The subsequent diffusion is a complex process, influenced by many factors such as temperature and local curvature of the graphene sheet.[28] Essentially, however, the atoms randomly interrogate a region of the surface until they encounter other Pt particles and nucleate into an island. As more Pt atoms arrive on the surface, the island grows. At a certain island size, covalent bonded anchors significantly alter the local bonding and the surface curvature due to the flexibility of the substrate. This in turn creates a local strain field in the immediate vicinity of the NP. Future Pt material therefore avoids these areas and preferentially migrates to other pristine regions of the surface. This process is somewhat similar to the self-organization of semiconductor quantum dot formation in strained thin film growth,[29] or to the diffusion of gallium (Ga) atoms under an arsenic (As) flux to form islands of a certain size and density on the surface of GaAs.[30] In our system, the Pt NPs interact with freestanding graphene in such a way as to repel each other over distances of a few nanometers and automatically limit the growth, resulting in uniform coverage across the sample but with a preferred size and nearest-neighbor separation.

**CONCLUSIONS**

In summary, freestanding graphene was uniformly functionalized with Pt NPs. HRTEM images revealed that the NPs have a narrow size distribution around 1.4 nm with preferential $(01\bar{1})$ surface orientation. Atomic-scale STM allowed characterization of the Pt NP surface topography as well as the binding properties to the graphene substrate. Through this unprecedented combination of advanced microscopy techniques, along with MD simulations, we were able to determine that the Pt NPs are not encapsulated by the graphene, and we also propose a strain mechanism for Pt NP self-organization. Since in general the size, shape, and orientation of Pt NPs are difficult to control, we see the potential to utilize freestanding graphene as a substrate which naturally controls these variables and simplifies the Pt NP growth process.[31] Our results indicate that the natural Pt NP arrangement on suspended graphene maximizes the surface area for catalytic activity by forming a somewhat flattened NP shape, extending the NPs into the vacuum rather than wrapping them, and distributing them with a uniform number density.

**METHODS**

*Sample Preparation.* Graphene was grown on a Ni foil *via* chemical vapor deposition,[32] transferred using isopropanol onto a 2000-mesh, ultrafine copper grid comprised of square holes 7.5 μm wide and bar supports 5 μm wide, and then the Ni substrate was etched away (Graphene Supermarket).[33] A continuous film having 1 to 6 monolayers is present, SEM images show 90% coverage, and grain size is estimated at 1-10 μm. Pt NPs were synthesized on the graphene surface *via* a simple, single-step physical sputtering process. Briefly, the copper grid with freestanding graphene was placed onto the sample holder of a desktop sputter coater (Ted Pella, 108 Auto Sputter Coater). Once the chamber was pumped down to 0.08 mbar, it was filled with Ar gas at a pressure of 0.1–0.15 mbar, and the sputtering deposition was conducted at 10 mA for 30 s. After venting the chamber to the atmosphere, the copper grid was removed from the chamber for characterization.

*HRTEM.* Since graphene was suspended on a 3 mm copper TEM grid, and the Pt NP deposition was conducted *via* a vacuum physical method, both the freestanding graphene and NP-functionalized graphene could be directly placed into a TEM sample holder without any chemical processes or sample preparations. The HRTEM experiments were conducted at 200 keV using a JEOL JEM-2010 microscope.

*STM imaging and electrostatic manipulation.* STM experiments were performed using an Omicron ultrahigh-vacuum, low-temperature model STM operated at room temperature. The samples were mounted on a flat tantalum sample plate using silver paint and loaded into the STM chamber *via* a load-lock. The STM tips were electrochemically etched from 0.25 mm diameter tungsten wire *via* a custom double-lamella setup with an automatic gravity-switch cutoff. After etching, the tips were gently rinsed with distilled water, briefly dipped in a concentrated hydrofluoric acid solution to remove surface oxides, and then loaded into the STM chamber. STS tip displacement data were obtained with the feedback electronics left on, meaning that the tunneling current was maintained at a constant setpoint.

*MD simulations.* The reactive force field ReaxFF[34, 35] was implemented in the large-scale atomic/molecular massively parallel simulator (LAMMPS) code.[36] A description of the force field appropriated for structures with carbon-Pt interaction can be found in Ref. [37]. The temperature was maintained by the Nosé-Hoover thermostat, and the MD time-step was taken to be 0.1 fs. A range of calculations were carried out for each simulation discussed, but the most relevant and representative results have been presented.

*Conflict of Interest:* The authors declare no competing financial interst.

*Supporting Information Available:* A movie in the supplementary material giving the self-assembling process of Pt Np on freestanding graphene. This material is available *via* the Internet at http://pubs.acs.org.

*Acknowledgements.* M.N.A. acknowledges financial support by the EU-Marie Curie IIF postdoc Fellowship/299855. F.M.P. acknowledges financial support by the ESF-EuroGRAPHENE project CONGRAN, the Flemish Science Foundation (FWO-Vl), and the Methusalem Foundation of the Flemish Government. L.D. acknowledges financial support by the Taishan Overseas Scholar program (tshw20091005), the International Science & Technology Cooperation Program of China (2014DFA60150), the National Natural Science Foundation of China (51172113), the Shandong Natural Science Foundation (JQ201118), the Qingdao Municipal Science and Technology Commission (12-1-4-136-hz), and the National Science Foundation (DMR-0821159). P.M.T. is thankful for the financial support of the Office of Naval Research under Grant No. N00014-10-1-0181 and the National Science Foundation under Grant No. DMR-0855358.

**Figure Captions**

**Figure 1.** (a) SEM image of pristine freestanding graphene supported by a Cu TEM grid. Within the holes of the grid the gray areas are graphene. (b) Zoomed-in SEM image showing graphene within the Cu bar supports as grey and lack of graphene as black. (c) Schematic showing the single-step sputtering process for deposition of Pt onto a freestanding graphene membrane.

**Figure 2.** (a-c) HRTEM images of Pt NPs supported by a freestanding graphene membrane with three different magnifications. Atomic-scale features, including the distances between atomic planes, can be resolved in (c). (d) Histogram showing particle size. (e) The ACF of the image in part (b). (f) A line profile taken along the line in part (e) and used to find an average particle size and nearest-neighbor distance.

**Figure 3.** (a) 12 nm × 12 nm filled-state STM image of Pt NPs on suspended graphene. The lines mark the locations where the line profiles were obtained. The horizontal line profile is plotted in blue directly below the image, and the vertical line profile is plotted in red below the first graph. Both show the height profile for the same NP. The lower left inset image is pristine graphene acquired further away from the NPs. The lower right inset image shows the artificially enlarge graphene at the spot marked by the white box. Both inset images are magnified and measure 1.2 nm by 1.2 nm. (b) Cropped image of the larger NP in part (a), displayed with a compressed color scale to show the atomic rows. A line profile was taken near the line shown and is plotted

directly below the image. Below that graph is a model of the Pt $(01\bar{1})$-oriented surface with the top layer highlighted.

**Figure 4.** (a) The initial arrangement of the Pt atoms over graphene in a MD simulation. (b) The final self-assembled structure after relaxation by MD computations at $T = 10$ K. A Pt NP was formed with a lateral size of 3 nm and height of 0.5 nm. See the corresponding movie in the supplementary material for the self-assembling process. The right (left) inset shows the side (top) view. (c) Five Pt NPs which were found to be stable at room temperature after a long MD simulation time when placed on the graphene surface.

**Figure 5.** (a) STM tip height *versus* bias voltage measurements taken at a setpoint current of 1.00 nA for the Pt-functionalized graphene (top curve), pristine freestanding graphene (middle curve), and inflexible Au (bottom curve). (b) Depiction of Pt atoms on graphene inside an initial Gaussian depression. (c) After MD relaxation the previous curvature in the graphene sheet disappeared, and the Pt NP resides on top of a flat graphene surface.

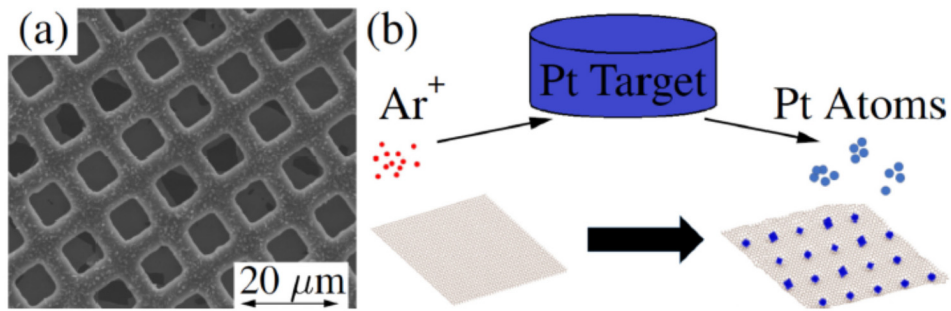

Figure 1

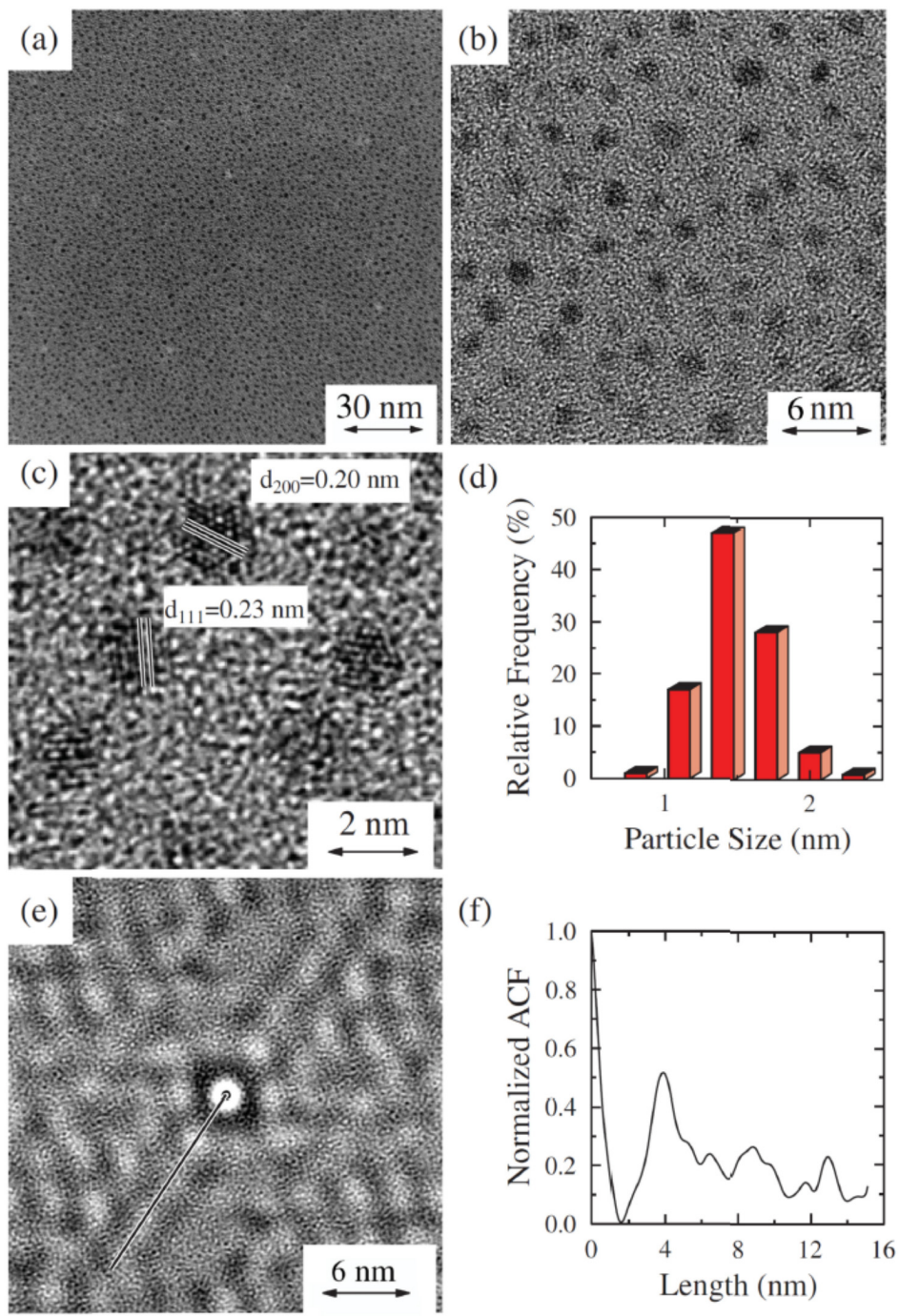

Figure 2

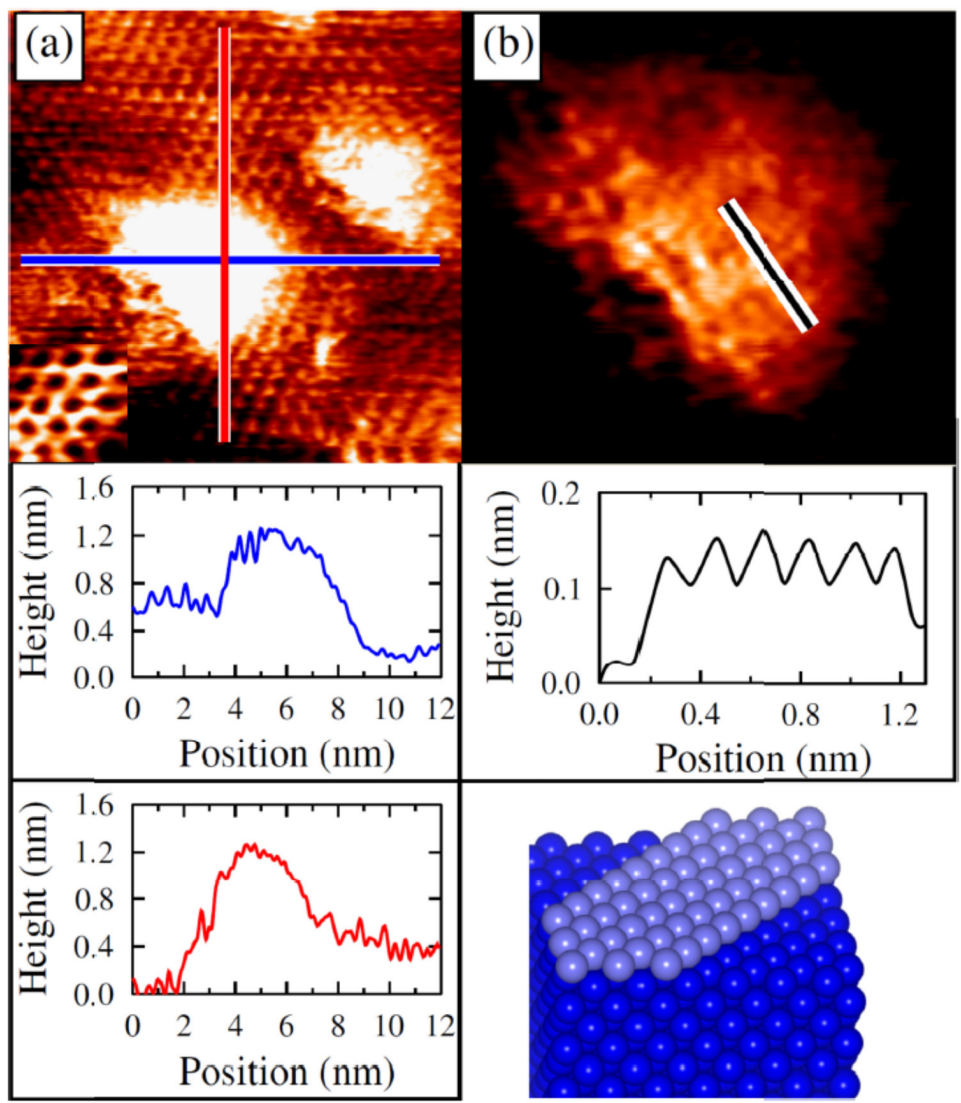

Figure 3

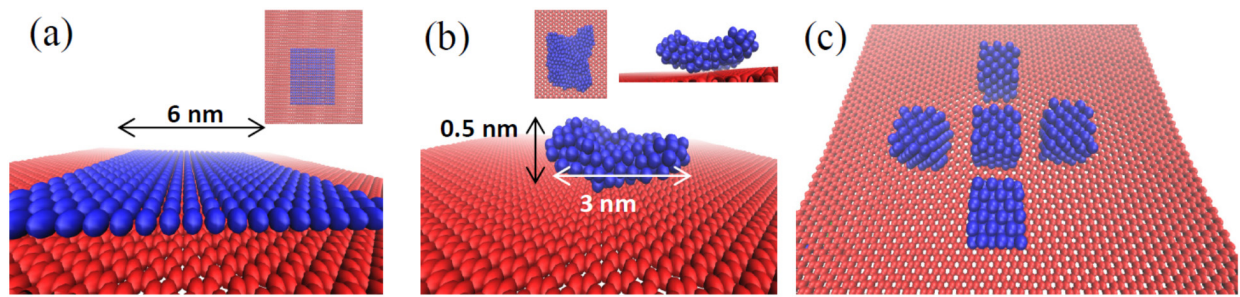

Figure 4

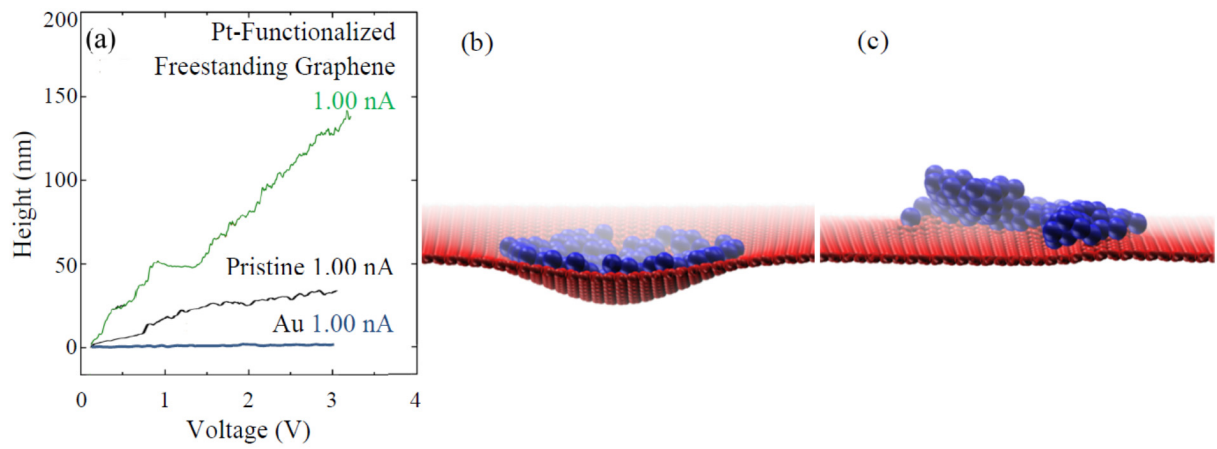

Figure 5